\documentclass[article,amsmath,amssymb]{revtex4-1}

\usepackage{tikz}
\begin{document}
\title{Brans-Dicke-Maxwell Solutions for Higher Dimensional Static Cylindrical Symmetric Spacetime}

\author{Dilek K. \c{C}iftci}
\email{dkazici@nku.edu.tr}
\affiliation{Physics Department, Nam\i k Kemal University, Tekirda\u g, Turkey}

\author{\"Ozg\"ur Delice}
\email{ozgur.delice@marmara.edu.tr}
\affiliation{Physics Department, Marmara University, Faculty of Science and Letters, 34722 Istanbul, Turkey}

\date{\today}
\begin{abstract}
In this paper, Brans-Dicke-Maxwell type vacuum solutions are considered for a static cylindrically symmetric spacetime in arbitrary dimensions. Exact solutions  are obtained by directly solving the field equations for the case where an azimuthal magnetic field is present. Other configurations such as axial magnetic field case can be obtained by suitably relabeling the coordinates. We have also considered conformally related "Einstein frame" to relate the solutions we have obtained with the dilaton-Maxwell type solutions that exist in the literature. We see that for a special case  the general solution we present reduces to Dilaton-Melvin spacetime.  The General Relativistic limit of these solutions are also discussed  and we found that this limit is different from the four dimensional case.   

\end{abstract}
\pacs{0.4}
\maketitle
\section{Introduction}
Brans-Dicke (BD) Scalar-Tensor theory \cite{BD-original} is the most studied alternative gravity theory, since it is well motivated and simple extension \cite{Faraonibook,fujui-maeda} of Einstein's General Relativity (GR). In this theory, Newton gravitational coupling constant is replaced by a scalar field nonminimally coupled to the gravitational action. The existence of this scalar is motivated by several theoretical results \cite{Faraonibook,fujui-maeda,Brans1}. For example, in the theories which aim to unify the  fundamental forces a la Kaluza-Klein type  compactification scheme, a scalar field called as the dilaton field naturally emerges. In order to understand any gravitational theory, obtaining and studying exact solutions of this theory, if possible, for different physically relevant configurations are mandatory.

Recently an interest is emerged in higher dimensional cylindrically symmetric solutions \cite{Clement,Bronnikov,Ponce,Bay-Oz-Di,Oz-Di-Pi}. This is motivated by, among others, the possible existence \cite{Tye,Tye2} of one dimensional extended bodies such as cosmic strings \cite{VilenkinCS} in the theories having higher dimensional spacetimes such as Brane-inflation theories. Also, it is known that stable configurations \cite{dfstrings,Copeland} of  macroscopic cosmic $F$ and $D$ superstrings \cite{Wittensuper} can be present. Hence, the gravitational field of an extended field configurations in higher dimensions must be investigated. Since cylindrically symmetric solutions can be a good approximation of such configurations, it is reasonable to obtain the exact solutions and investigate their further properties of cylindrically symmetric solutions in GR and its alternative theories such as BD theory.

In four dimensional space-time, cylindrically symmetric exact vacuum \cite{LC} and Maxwell vacuum solutions \cite{Bonnor,Raychaudhuri}  were presented decades ago and  their further properties were investigated in different contexts \cite{Melvin,LWitten,Safko,MacCallum,Richterek,Miguelote,BaykalEM}. These solutions are especially useful to understand the exterior gravitational field of straight cosmic \cite{VilenkinCS} and superconducting cosmic strings \cite{Superconducting,Moss,PeterPuy}  which produce electric and magnetic fields around them. Extending these kind of configurations to higher dimensions, one needs the exact vacuum or Maxwell vacuum solutions in a higher dimensional spacetime. Since static vacuum \cite{Ponce}, Einstein-Maxwell vacuum {\cite{Oz-Di-Pi} solutions  in GR and vacuum solutions in BD theory for four \cite{Ar-Si} and higher dimensions \cite{Bay-Oz-Di} were already presented, in this paper we investigate Maxwell vacuum solutions in BD theory for higher dimensions. The paper is organized as follows. In Sec. II,  the field equations of the BD-Maxwell action for a general static cylindrically symmetric spacetime will be presented in the presence of a magnetic field along the axial direction. In section III, their exact solutions will be presented and their various limits will be discussed. Section V is devoted to the discussion of GR limit of these solutions. In Section V, the corresponding solutions in Dilaton-Maxwell gravity will  be obtained by a conformal transformation. The paper ends with some concluding comments.

\section{Brans-Dicke-Maxwell Action and the SpaceTime}

We consider Brans-Dicke Scalar-Tensor theory in $n$ spacetime dimensions in the presence of a Maxwell field as described by the following action in the Jordan frame
\begin{equation}
S=\int d^nx\sqrt{-g}\left(\Phi R-\frac{\omega}{\Phi}\partial_{\mu}\Phi\partial^{\mu}\Phi-F_{\mu\nu}F^{\mu\nu}\right)\label{action},
\end{equation}
where the scalar field $\Phi^{-1}$ replaces  the Newton's gravitational constant in the Einstein-Hilbert action. The dimensionless parameter $\omega$ is the BD coupling constant. The Faraday two form  $F_{\mu\nu}$ is coupled minimally to the action.  The field equations are obtained by varying the action with respect to the metric $g_{\mu\nu}$,
\begin{equation}
G_{\mu\nu}=\frac{\omega}{\Phi^2}\left( \partial_{\mu}\Phi\partial_{\nu}\Phi-\frac{1}{2}g_{\mu\nu}\partial_{\alpha}\Phi\partial^{\alpha}\Phi\right)+\frac{1}{\Phi}\left( \partial_{\mu}\partial_{\nu}\Phi-g_{\mu\nu}\partial_{\alpha}\partial^{\alpha}\Phi\right)+\frac{2}{\Phi}\left(F_{\mu\lambda}F^{\lambda}_{\ \nu}-\frac{1}{4}g_{\mu\nu}F_{\alpha\lambda}F^{\alpha\lambda} \right)  \label{varg},
\end{equation}
and the varying with respect to the scalar field $\Phi$ yields,
\begin{equation}
R=\frac{\omega}{\Phi^2}\partial_{\mu}\Phi\partial^{\mu}\Phi-\frac{2\omega}{\Phi}\square\Phi \label{varphi}.
\end{equation}
Contracting (\ref{varg}) with the inverse metric $g^{\mu\nu}$ and summing with (\ref{varphi}) we have
\begin{equation}
\square\Phi=-\frac{n-4}{2\left[\omega (n-2)+(n-1) \right] }F_{\mu\nu}F^{\mu\nu}\label{box}.
\end{equation}
Here it is clear that the electromagnetic field is the source of scalar field $\Phi$  for $n>4$.
Note that the Maxwell equations
\begin{equation}
 dF=0, \quad d*F=0 ,
\end{equation}
must also be satisfied where $F=\frac{1}{2}F_{\mu\nu}dx^\mu \wedge dx^\nu$ and * is the Hodge star operator.

For the spacetime  metric, similar to the previous works \cite{Ponce,Bay-Oz-Di,Oz-Di-Pi}, we restrict ourselves with an $n$ dimensional static cylindrically symmetric spacetime  having $n-1$ commuting orthogonal Killing vectors $\partial/\partial_t,\ \partial/\partial_z,\ \partial/\partial_\phi, \ \partial/\partial x_i $. Hence the spacetime metric we consider is given by
\begin{equation}
ds^2=e^{2\left(K-U\right)}\left(-dt^2+dr^2 \right)+e^{2U}dz^2+e^{-2U}W^2d\phi^2+ \sum_{i=5}^{n}X_i^2(dx^i)^2,\label{metric3}
\end{equation}
where $i=5,...,n$, the coordinates $t,r,z,\phi$ define four dimensional cylindrical coordinates and $x^i$ denotes the extra coordinates. The metric functions, $K,\ U,\ W$ and $X_i,$ depend on the radial coordinate $r$ only. The BD type solutions of  more general higher dimensional axially symmetric spacetimes\cite{Emparan,Harmark} without a translational isometry along $z$ direction will not be discussed in this paper. We consider the electromagnetic potential one form as
\begin{equation}
A=f(r)dz,\label{elpot}
\end{equation}
then the magnetic field $F=f'dr\wedge dz$ becomes azimuthal.

Now let us investigate the field equations of above configuration.
We rename the right hand side of (\ref{varg}) with $\bar{T}_{\mu\nu}$ and write this equation as $G_{\mu \nu}-\bar{T}_{\mu\nu}=0$. Here $\bar{T}_{\mu\nu}$ includes contribution from the  scalar potential and electromagnetic fields.  The nontrivial components of the metric equations of Brans-Dicke-Maxwell (BDM) equations become,
\begin{eqnarray}
G^t_{\ t}-\bar{T}^t_{\ t}=&&e^{2(U-K)}\Bigg\lbrace U'^2-\frac{K'W'}{W}+\frac{W''}{W}+\sum_i^N\left[\left( U'-K'+\frac{W'}{W} +\frac{1}{2}\sum_{j\neq i}^N\frac{X'_j}{X_j}\right) \frac{X'_i}{X_i}+\frac{X''_i}{X_i}\right] \nonumber \\&&-\left( -\frac{e^{-2U} f'^2}{\Phi}-\frac{U'\Phi'}{\Phi}+\frac{K'\Phi'}{\Phi}-\frac{W'\Phi'}{W\Phi}-\frac{\omega}{2}\frac{\Phi'^2}{\Phi^2}-\frac{\Phi''}{\Phi}-\sum_{i}^{N}\frac{\Phi'}{\Phi}\frac{X_i'}{X_i}\right )\Bigg\rbrace=0, \label{kkG} \\
G^r_{\ r}-\bar{T}^r_{\ r}=&&e^{2(U-K)}\Bigg\lbrace -U'^2+\frac{K'W'}{W}+\sum_i^N\left[\left(- U'+K'+\frac{W'}{W} +\frac{1}{2}\sum_{j\neq i}^N\frac{X'_j}{X_j}\right) \frac{X'_i}{X_i}\right]\nonumber \\&&-\left( \frac{e^{-2U} f'^2}{\Phi}+\frac{U'\Phi'}{\Phi}-\frac{K'\Phi'}{\Phi}-\frac{W'\Phi'}{W\Phi}+\frac{\omega}{2}\frac{\Phi'^2}{\Phi^2}-\sum_{i}^{N}\frac{\Phi'}{\Phi}\frac{X_i'}{X_i}\right )\Bigg\rbrace=0, \label{rrG}
\\
G^z_{\ z}-\bar{T}^z_{\ z}=&&e^{2(U-K)}\Bigg\lbrace U'^2-2U''+K''+\frac{W''}{W}-\frac{2U'W'}{W}+\sum_i^N\left[\left( -U'+\frac{W'}{W} +\frac{1}{2}\sum_{j\neq i}^N\frac{X'_j}{X_j}\right) \frac{X'_i}{X_i}+\frac{X''_i}{X_i}\right]\nonumber  \\&&-\left( \frac{e^{-2U} f'^2}{\Phi}+\frac{U'\Phi'}{\Phi}-\frac{W'\Phi'}{W\Phi}-\frac{\omega}{2}\frac{\Phi'^2}{\Phi^2}-\frac{\Phi''}{\Phi}-\sum_{i}^{N}\frac{\Phi'}{\Phi}\frac{X_i'}{X_i}\right )\Bigg\rbrace=0 ,\label{zzG}\\
G^{\phi}_{\ \phi}-\bar{T}^\phi_{\ \phi}=&&e^{2(U-K)}\Bigg\lbrace U'^2+K''+\sum_i^N\left[\left( U'+\frac{1}{2}\sum_{j\neq i}^N\frac{X'_j}{X_j}\right) \frac{X'_i}{X_i}+\frac{X''_i}{X_i}\right]\nonumber \\&&-\left(- \frac{e^{-2U} f'^2}{\Phi}-\frac{U'\Phi'}{\Phi}-\frac{\omega}{2}\frac{\Phi'^2}{\Phi^2}-\frac{\Phi''}{\Phi}-\sum_{i}^{N}\frac{\Phi'}{\Phi}\frac{X_i'}{X_i}\right )\Bigg\rbrace=0, \label{ppG}\\
G^{x_i}_{\ x_i}-\bar{T}^{x_i}_{\ x_i}=&&e^{2(U-K)}\Bigg\lbrace U'^2-U''-\frac{U'W'}{W}+K''+\frac{W''}{W}+\sum_{j\neq i}\left[\left(  \frac{W'}{W}+\frac{1}{2}\sum_{k\neq j\neq i}\frac{X'_k}{X_k} \right) \frac{X'_j}{X_j}+\frac{X_j''}{X_j}\right] \nonumber\\&&-\left( -\frac{e^{-2U} f'^2}{\Phi}-\frac{W'\Phi'}{W\Phi}-\frac{\omega}{2}\frac{\Phi'^2}{\Phi^2}-\frac{\Phi''}{\Phi}-\sum_{j\neq i}\frac{\Phi'}{\Phi}\frac{X_j'}{X_j}\right )\Bigg\rbrace=0, \label{iiG}
\end{eqnarray}
where we have define the abbreviations  $X=\prod_{i=4}^{D-1}X_i$ and $\Omega=\Phi W X$.
From (\ref{kkG}) and (\ref{rrG}) we have
\begin{equation}(\Omega)''=0,
 \end{equation}
which implies the important relation
\begin{equation}
 \Omega= \Omega_0 r. \label{comp1}
\end{equation}

From (\ref{kkG}) and (\ref{ppG}) we find
\begin{equation}\left( \Omega K'\right) '-\left(  \Omega\frac{W'}{W}\right) '=0,
\end{equation}
which can be easily integrated as
\begin{equation}
  K'-\frac{W'}{W}=\frac{c_2}{r}.
\end{equation}
Moreover using (\ref{kkG}) and (\ref{iiG}) we obtain
\begin{equation}
-\left(\Omega K' \right)'+\left( \Omega U'\right)'+\left( \Omega\frac{X_i'}{X_i}\right)'=0,
\end{equation}
which can also be integrated to get
\begin{equation}
  -K'+U'+\frac{X'_i}{X_i}=\frac{c_4}{r}.
\end{equation}
One other usefull equation is, using (\ref{ppG}) and (\ref{iiG}) we find
\begin{equation}
-\left(\Omega U' \right)'+\left( \Omega \frac{W'}{W}\right)'-\left( \Omega\frac{X_i'}{X_i}\right)'=0,
\end{equation}
which can also be integrated to find
\begin{equation}
-U'+\frac{W'}{W}-\frac{X_i'}{X_i}=\frac{c_4}{r}.
\end{equation}
Moreover by picking from the equations (\ref{iiG}) two different expressions for the values of the indices $i,j=5,6,7...$  and adding them
yields
\begin{equation}
\left( \Omega\frac{X'_i}{X_i}\right)'-\left( \Omega\frac{X'_j}{X_j}\right)'=0,
\end{equation}
whose straightforward integration results the following first order differential equation
\begin{equation}
 \frac{X'_i}{X_i}-\frac{X'_j}{X_j}=\frac{c_{ij}}{r}. \label{comp5}
  \end{equation}

From the Maxwell equations $d*F=0$, we have found that
\begin{equation}
f'=f_0\frac{e^{2U}}{WX }\label{Maxwell}.
\end{equation}
Here $f_0$ is an integration constant and we define the  $X=\prod_{i=4}^{D-1}X_i$ for  metric functions of extra dimensions.
Moreover from the scalar field equation (\ref{box}) we find
\begin{equation}
\frac{e^{2U}}{WX}\left( \Phi'WX\right) '=-\frac{n-4}{ (n-1)+(n-2)\omega  }f'^2 .\label{box2}
\end{equation}
These two equations give the following relation:
\begin{equation}
\Phi'WX =-\frac{n-4}{\left(n-1 \right)+\left(n-2 \right)\omega  }f_0 f+f_1, \label{fr2}
\end{equation}
where $f_1$ is an integration constant. Here it is easily seen that  for $n=4$ or vacuum cases, the trace of energy momentum tensor of electromagnetic field vanishes, therefore the first term on the right hand side gives no contribution.

By algebraically manipulating all of the first order differential equations given above can reduce to a a single differential equation for one of the metric functions, whose integration yields the desired solution. Skipping the details, the solution of BDM field equations with suitably chosen integration constants will be presented in Sec. III.

\section{Solutions}

 Here we present the results as follows
\begin{eqnarray}
&&\Phi=\phi_0 r^{1-k}(1+c^2r^p)^{-\frac{2(n-4)}{\omega_n}},\label{phi}\\
&&W=W_0 r^{k-l}(1+c^2r^p)^{-\frac{n-4}{n-3}\left(1-\frac{n-2}{\omega_n} \right) },\\
&&K=q \ln r-\frac{n-4}{n-3}\left(1-\frac{n-2}{\omega_n} \right) \ln(1+c^2r^p),\\
&&U=d \ln r-\left(1- \frac{n-4}{\omega_n}\right)  \ln(1+c^2r^p),\\
&&X_i=X_{i0} r^{l_i}(1+c^2r^p)^{\frac{1}{n-3}\left(1+\frac{n-4}{\omega_n} \right) },\label{Xi}\\
&&f=\pm\sqrt{\frac{n-2}{ n-3} -\frac{\left(n-4 \right)^2 }{\omega_n\left( n-3\right) }}\frac{\sqrt{\Phi_0}}{\sqrt{2}c(1+c^2r^p)}+f_2 . \label{fr}
\end{eqnarray}
Here $\phi_0, W_0,\Phi_0, f_2$ and the constant parameters given below are integration constants:
\begin{eqnarray}
&&\omega_n= 4\left(n-3 \right)\left(\omega+1 \right)+\left(n-2 \right)    \label{pi}\\
&&l=\sum_{i=4}^{n-1}l_i \\
&&m^2=\sum_{i=4}^{n-1}l_i^2\\
&&p=2d+1-k\\
&&q=d(d+1-k+l)+\frac{\omega}{2}(1-k)^2-k(1-k+l)+\frac{l^2+m^2}{2}. \label{q}
\end{eqnarray}
Note that this solution is invariant under the transformation   $p\rightarrow -p$.
All these metric functions (\ref{phi}-\ref{Xi}) and constants (\ref{pi}-\ref{q}) satisfy the BDM field equations and magnetic field becomes
\begin{equation}
B_{\phi}=f'=\pm \frac{\sqrt{\phi_0}}{\sqrt{2}}\sqrt{\frac{n-2}{ n-3} -\frac{\left(n-4 \right)^2 }{\omega_n\left( n-3\right) }}\frac{c p\ r^{p-1}}{(1+c^2r^p)^2}. \label{fr1}
\end{equation}

Note also that for the sake of the consistency of the equations (\ref{fr}) and (\ref{fr2}), the integration constant $f_0$  given in (\ref{Maxwell}) must be $f_0=\mp \frac{cp\sqrt{\phi_0}}{\sqrt{2}}W_0X_0 \sqrt{\frac{n-2}{ n-3} -\frac{\left(n-4 \right)^2 }{\omega_n\left( n-3\right) }}$ and $k$ becomes
\begin{equation}
k=1-\frac{2p\left(n-4\right) }{\omega_n}\label{k}.
\end{equation}
This relation is only present in $n>4$ and in four dimensions the parameter $k$ is arbitrary. This is a crucial result in discussing the GR limit of the solutions we present.

The line element of the solution can be expressed in a more compact form,
\begin{equation}
ds^2=G(r)^{\frac{2}{n-3}} \left[r^{2(q-d)}\left(-dt^2+dr^2 \right)+W_0^2r^{2(k-d-l)}d\phi^2+\sum X_{i0}^2r^{2l_i}dx_i^2  \right] +\left[ G(r)^{\frac{n-4-\omega_n}{n-4+\omega_n}} r^{d}\right] ^2dz^2,  \label{metric1}
\end{equation}
where $G(r)=(1+c^2r^p)^{\frac{n-4}{\omega_n}+1}$. Here it can be easily seen that for the four dimensional case (\ref{metric1}) reduces the results in \cite{BaykalEM}.

In order to obtain BD vacuum solution, one needs to set  $c=0$, which yields the magnetic field (\ref{fr1}) to be vanish. Then line element is given by
\begin{equation}
ds^2=r^{2(q-d)}\left(-dt^2+dr^2 \right)+r^{2d}dz^2+W_0^2r^{2(k-d-l)}d\phi^2+\sum X_{i0}^2r^{2l_i}dx_i^2,
\end{equation}
and the BD scalar  becomes $\Phi=\Phi_0r^{1-k}$. This solution corresponds to  static cylindrically symmetric vacuum BD solution \cite{Bay-Oz-Di} in Einstein-Rosen type coordinates in arbitrary dimensions.
For the $n=4$ and $c=0$,  four dimensional BD vacuum solution given in \cite{Ar-Si}  is obtained as
\begin{equation}
ds^2=r^{2(q-d)}\left(-dt^2+dr^2 \right)+r^{2d}dz^2+W_0^2r^{2(k-d)}d\phi^2,
\end{equation}
with $\Phi=\Phi_0r^{1-k}$. For the limit $k=1$, scalar field becomes constant, thus Levi-Civita GR vacuum metric is obtained \cite{LC},
\begin{equation}
ds^2=r^{2d(d-1)}\left(-dt^2+dr^2 \right)+r^{2d}dz^2+W_0^2r^{2(1-d)}d\phi^2 .
\end{equation}

\textit{For $c\neq 0$}, there is an electromagnetic field distribution in the space time.

Note that, following a similar strategy for  the solution above, obtaining a different Electromagnetic  field configuration is not difficult. For example
if there is a current in the azimuthal angular direction, we have a solution in the general form of  (\ref{metric1}) which can either be obtained by solving the field equations or suitably relabeling the coordinates and the parameters. Namely, for an electromagnetic potential one form $A(r)=f(r)d\phi$, the Maxwell equations are satisfied for $f(r)'=f_0 e^{-2U}\frac{W}{X}$ and BDM equations describe the space time as
\begin{equation}
ds^2=G(r)^{\frac{2}{n-3}} \left[r^{2(q-d)}\left(-dt^2+dr^2 \right)+r^{2d}dz^2+\sum X_{i0}^2r^{2l_i}dx_i^2  \right] +\left[ G(r)^{\frac{n-4-\omega_n}{n-4+\omega_n}} r^{k-d-l}\right] ^2d\phi^2,     \label{metric5}
\end{equation}
and the current density becomes $f=\pm\frac{\sqrt{\Phi_0}}{\sqrt{2}}\sqrt{\frac{n-2}{ n-3} -\frac{\left(n-4 \right)^2 }{\omega_n\left( n-3\right) }}\frac{1}{c(1+c^2r^p)}+f_2$.
Note that if one seeks a solution with a radial electrical field, Einstein-Rosen type coordinates are not adequate and one needs to use Weyl type metric.

These magneto-vacuum solutions have a  genuine curvature singularity at $r=0$ which can be easily seen by investigating curvature  and field scalars. For example, the  Lorentz invariant and Ricci scalar  become,
\begin{eqnarray}
&&F_{\mu\nu}F^{\mu\nu}=\phi_0\left[ \frac{n-2}{n-3}-\frac{\left(n-4 \right)^2 }{\omega_n\left(n-3 \right) }\right] p^2c^2r^{-2(1+q-p)}\left(1+c^2r^p \right)^{-2\frac{n-2}{n-3}\left(1+ \frac{n-4}{\omega_n}\right) }=2B^2>0\label{lrntz}\\
&&R=\frac{n-4}{n-3  }\left[ 1-\frac{5n-14}{\omega_n}\right] p^2r^{-2(1+q-d)}\left(1+c^2r^p \right)^{-2\frac{n-2}{n-3}\left( 1+\frac{n-4}{\omega_n\left(n-2 \right) }\right) } \left( \frac{n-4}{\omega_n}+c^2r^p\right) .
\end{eqnarray}
The other scalars such as  Ricci and Kretschmann scalars indicate same values for singularities. From (\ref{lrntz}), we note that, since $p$ is a real parameter, $c^2$ must be always positive. This implies the result that there is only one singularity at $r=0$ and this singularity is unavoidable unless the parameters are chosen such that the exponents of the term $r$ outside the parentheses vanishing.

\section{General Relativistic Limit}
The obtained relation (\ref{k}) between Brans-Dicke parameter $\omega$ and the extra constant $k$ permits us to find the General Relativistic limit easily. It is known that the $\omega\rightarrow \infty$  limit do not always work \cite{Matsuta,romero-barros1,Paiva1,Paiva2,romero-barros,Scheel,banerjee-sen,Anchordoqui,faraoni1,faraoni,faraoni2},   as the correct GR limit of BD theory when one considers a vacuum solution or a  non-vacuum solution with a traceless energy momentum tensor. For this cases, usually, correct  limit is obtained by setting arbitrary constants of the solutions related to scalar field to specific values, unless the arbitrary constants of the solutions can be related to the parameter $\omega$ \cite{bhadra-nandi} by some other mechanism, such as matching with a regular interior solution in the Post-Newtonian expansion for spherically symmetric vacuum solutions \cite{BhadraSarkar} .

Thus for $n>4$ and for the limit $\omega\rightarrow\infty$, this solution reduces the GR one \cite{Oz-Di-Pi} and $k$ becomes  unity, since the scalar field becomes constant which may be identified as Newton's gravitational constant $G_N=1/\Phi_0$ . Here, the crucial thing is the equation (\ref{k})  does not exist for  four dimensional or vacuum spacetimes because of the condition in (\ref{fr2}). Namely, for $n=4$ or vacuum case,  $k$ becomes a free parameter and independent of $\omega$ and GR limit is obtained by setting  $k=1$ \cite{BaykalEM}. The equation (\ref{k})  appears only for Brans-Dicke-Maxwell solutions of higher dimensional ($n>4$) cylindrically symmetric spacetime. Therefore, this solution is a good example to distinguish the cases where   $\omega \rightarrow \infty$ is the correct GR limit or not. Namely, in our solution, in four dimensions, unlike $d>4$, the Einstein limit is not $\omega\rightarrow\infty$, but $k=1$. This is due to the fact that in four dimensions, the trace of the Maxwell energy momentum tensor is vanishing. In $d>4$, however, it does not vanish, and the correct limit becomes $\omega\rightarrow\infty$.

\section{Solutions in Einstein Maxwell Dilaton Theory}

It is a well known fact that  by a conformal transformation of the form
\begin{eqnarray}
&&\tilde{g}_{\mu\nu}=\Phi^{\frac{2}{n-2}}g_{\mu\nu},\\
&&\Psi= \sqrt{\frac{1}{2}\left(\omega+\frac{n-1}{n-2} \right)}\, \ln \Phi,
\end{eqnarray}
BD action (\ref{action}), expressed in the Jordan frame, can be put in the ``Einstein frame'' as follows
\begin{equation}
S=\int d^nx\sqrt{-\tilde g}\left( \tilde R-2\partial_{\mu}\Psi\partial^{\mu}\Psi-e^{-2\alpha \Psi}F_{\mu\nu}F^{\mu\nu}\right), \label{actiondil}
\end{equation}
where the relation between the Brans Dicke parameter $\omega$ and dilaton coupling constant $\alpha$ is given by
 \begin{equation}
\omega=\frac{1}{2}\left[\frac{n-4}{\alpha\left(n-2 \right)}\right]^2-\frac{n-1}{n-2}.
 \end{equation}
Here we use $\tilde {}$ to denote either tensorial quantities calculated from the transformed metric $\tilde g_{\mu\nu}$ or the  constants of the solutions of the field equations of this action.
In this form of the action, although the new scalar field, the dilaton field $\Psi$, is not coupled to gravity directly, it is nonminimally coupled to the Maxwell term in the action. Due to this fact, test particles do not follow geodesics of the metric $\tilde g_{\mu\nu}$.
The field equations of the action (\ref{actiondil}), which are also  called as Einstein-Maxwell-dilaton  (EMd) equations,  are given by
\begin{eqnarray}
&&\tilde G_{\mu\nu}=2\left(\tilde \nabla_{\mu}\Psi \tilde \nabla_{\nu}\Psi-\frac{1}{2} \tilde g_{\mu\nu} \tilde \nabla_{\alpha} \Psi \tilde \nabla^{\alpha}\Psi\right)+2 e^{-2 \Psi\alpha}\left(F_{\mu\lambda}F^{\lambda}_{\ \nu}-\frac{1}{4} \tilde g_{\mu\nu} F_{\beta\lambda}F^{\beta\lambda} \right),\\
&&\tilde \square\Psi=-\frac{\alpha}{2}e^{-2\alpha\Psi}F_{\beta\lambda}F^{\beta\lambda},\\
&&\tilde \nabla_{\mu}\left[e^{-2\alpha\Psi}F^{\mu\nu} \right]=\tilde \nabla_{[\mu }F_{\nu\lambda]} =0.
\end{eqnarray}

Either applying the conformal transformation given above or directly solving the field equations of the action (\ref{actiondil}) using a similar ansatz for the metric (\ref{metric3}) and Maxwell field (\ref{elpot}), the general solution of EMd theory for $n$ dimensional static cylindrically symmetric spacetime is given by
\begin{equation}
ds^2=\tilde{G}(r)^\frac{2}{n-3}\left[r^{2(\tilde{q}-\tilde{d})}\left(-dt^2+dr^2 \right)+W_0^2r^{2(1-\tilde{d}-\tilde{l})}d\phi^2+\sum X_{i0}^2r^{2\tilde{l}_i}dx_i^2  \right] +\tilde{G}\left(r \right) ^{-2} r^{2\tilde{d}}dz^2,   \label{dilaton}
\end{equation}
where $\tilde{G}(r)=(1+c^2r^p)^{\frac{1}{1+\alpha^2_n}}$, and $\alpha_n=\sqrt{\frac{n-2}{2(n-3)}}\alpha$. The dilaton field satisfies
\begin{equation}
e^{-2\alpha\Psi}=r^{-2\alpha\psi_1}(1+c^2r^p)^{\frac{2\alpha^2_n}{1+\alpha^2_n}}\label{potdilaton},
\end{equation}
and the current density becomes
\begin{equation}
f=\pm\sqrt{\frac{n-2}{2(n-3)}}\frac{1}{\sqrt{1+\alpha^2_n}}\frac{1}{c(1+c^{\pm 2}r^{\pm p})}+f_0.\label{fdilaton}
\end{equation}
The other integration constants are obtained as,
\begin{eqnarray}
&&\tilde{l}=\sum_{i=4}^{n-1}\tilde{l}_i \\
&&\tilde{m}^2=\sum_{i=4}^{n-1}\tilde{l}_i^2\\
&&\tilde{d}=\frac{1}{2}p-\alpha \psi_1\\
&&\tilde{q}=\tilde{d}^2+\tilde{d}\ \tilde{l}-\tilde{l}+\psi_1^2+\frac{\tilde{l}^2+\tilde{m}^2}{2}. \label{qd}
\end{eqnarray}

As far as we know, the only solution of either BDM or EMd theories for higher dimensional cylindrical spacetime with a Maxwell field present is the
Dilaton-Melvin solution presented in \cite{Gibbons}, see also \cite{Yazadjiev}. This solution is obtained by using a Harrisson transformation \cite{Harrison} for dilaton gravity \cite{Dowker,Yazadjiev}.  We now show that this solution is a special case of the general solution we have presented above in (\ref{dilaton},\ref{potdilaton},\ref{fdilaton}). Setting the integration constants to specific values as $\tilde{l}_i=0$ and $\tilde{d}=1$ yields $\tilde{q}=1$, $p=2$ and $\psi_1=0$. Therefore the above solution reduces to
\begin{eqnarray}
&ds^2=\tilde{G}(r)^\frac{2}{n-3}\left[-dt^2+dr^2+W_0^2d\phi^2+\sum X^2_{i0}dx_i^2  \right] +\tilde{G}\left(r \right) ^{-2} r^{2}dz^2,\\
&f= \pm\sqrt{\frac{n-2}{2(n-3)}}\frac{1}{\sqrt{1+\alpha^2_n}}\frac{1}{c(1+c^{ 2}r^{ 2})}+f_0.\label{fmelvin}
\end{eqnarray}
This solution is nothing but Dilaton-Melvin solution if we interchange the coordinates $z$ and $\phi$ as well.
\section{Conclusion}

In this work, we have presented the BDM solutions for higher dimensional static cylindrically symmetric space time. In the main part of the paper, we have considered the "$n$" current along the $z$ axis and all the metric  functions depend only on the radial coordinate. We have also presented  the solutions in the Einstein frame and showed that our solution contains higher dimensional Dilaton-Melvin solution as a special case. The general relativistic limit of our solutions have also been discussed and  we have seen that $\omega\rightarrow \infty$ limit works for higher dimensional case but not in four dimensions where the energy-momentum tensor is traceless. Using a simple relabeling of coordinates, other field configurations such as  a magnetic field along the symmetry axis or a radial magnetic field  can be derived straightforwardly.

\end{document}